\documentclass[conference]{IEEEtran}
\IEEEoverridecommandlockouts
\usepackage{cite}
\usepackage{amsmath,amssymb,amsfonts}
\usepackage{algorithmic}
\usepackage{graphicx}
\usepackage{textcomp}
\usepackage{xcolor}
\def\BibTeX{{\rm B\kern-.05em{\sc i\kern-.025em b}\kern-.08em
    T\kern-.1667em\lower.7ex\hbox{E}\kern-.125emX}}
\begin{document}

\title{Trans2Unet: Neural fusion for Nuclei Semantic Segmentation\\}

\author{\IEEEauthorblockN{Dinh-Phu Tran}
\IEEEauthorblockA{\textit{Department of Automation Engineering} \\ 
\textit{School of Electrical and Electronic Engineering} \\
\textit{Hanoi University of Science and Technology} \\
Hanoi, Vietnam \\
phu.td181692@sis.hust.edu.vn}
\and
\IEEEauthorblockN{Quoc-Anh Nguyen}
\IEEEauthorblockA{\textit{Department of Automation Engineering} \\ 
\textit{School of Electrical and Electronic Engineering} \\
\textit{Hanoi University of Science and Technology} \\
Hanoi, Vietnam \\
anh.nq181325@sis.hust.edu.vn}
\and
\IEEEauthorblockN{Van-Truong Pham}
\IEEEauthorblockA{\textit{Department of Automation Engineering} \\ 
\textit{School of Electrical and Electronic Engineering} \\
\textit{Hanoi University of Science and Technology} \\
Hanoi, Vietnam \\
truong.phamvan@hust.edu.vn}
\and
\IEEEauthorblockN{Thi-Thao Tran$^*$}
\IEEEauthorblockA{\textit{Department of Automation Engineering} \\ 
\textit{School of Electrical and Electronic Engineering} \\
\textit{Hanoi University of Science and Technology} \\
Hanoi, Vietnam \\
thao.tranthi@hust.edu.vn}}

\maketitle

\begin{abstract}

Nuclei segmentation, despite its fundamental role in histopathological image analysis, is still a challenge work. The main challenge of this task is the existence of overlapping areas, which makes separating independent nuclei more complicated. In this paper, we propose a new two-branch architecture by combining the Unet and TransUnet networks for nuclei segmentation task. In the proposed architecture, namely Trans2Unet, the input image is first sent into the Unet branch whose the last convolution layer is removed. This branch makes the network combine features from different spatial regions of the input image and localizes more precisely the regions of interest. The input image is also fed into the second branch. In the second branch, which is called TransUnet branch, the input image will be divided into patches of images. With Vision transformer (ViT) in architecture, TransUnet can serve as a powerful encoder for medical image segmentation tasks and enhance image details by recovering localized spatial information. To boost up Trans2Unet efficiency and performance, we proposed to infuse TransUnet with a computational-efficient variation called ”Waterfall” Atrous Spatial Pooling with Skip Connection (WASP-KC) module, which is inspired by the ”Waterfall” Atrous Spatial Pooling (WASP) module. Experiment results on the 2018 Data Science Bowl benchmark show the effectiveness and performance of the proposed architecture while compared with previous segmentation models.
%Note: trong abstract không được đưa trích dẫn

\end{abstract}


\section{Introduction}

Cell nuclei segmentation has been a crucial problem that attracts interest because of its practical applications in the diagnosis of cancer. In general, this task is similar to natural image segmentation, which involves the process of extracting desired objects from a nuclei image (image 2D or 3D), and can be done manually, semi-automatically, or full-automatically \cite{b5}\cite{b6}\cite{b7}. Recently, many deep learning models with high accuracy have been used for nuclei segmentation \cite{b8}. In 2015, Unet featured an encoder-decoder architecture combined with skip-connections to retain important features has showed outstanding results for segmentation tasks, especially medical images.\\
Although having been powerful network architectures, Unet and other CNN networks in general still have limitations in reproducing straightforward long-range interrelationships resulted from the intrinsic locality of convolution operations. Unlike the CNN-based networks, the models based on Transformers have global computing features. In \cite{b2}, TransUnet was proposed to solve that problem by employing a hybrid CNN-Transformer approach to enhance both elaborate high-resolution spatial information from feature maps of CNN and the global context, which is encoded by Transformers. Although Transformer has gained popularity in Computer Vision due to global features, the lack of low-level details makes local feature information extraction insufficient \cite{b10}.\\
To take full advantages of Unet and TransUnet, in this study, we propose to combine these two architectures to obtain a more powerful architecture. The proposed architecture named as Trans2Unet includes two main branches. One branch sends the input image through the Unet network, the other branch sends the input image through the TransUnet network. Finally, the outputs of these branches are concatenated to recreate feature maps of the input image, thereby improving the predictive ability of the model. Furthermore, instead of using the original TransUnet architecture, we added the WASP-KC module to leverage the progressive extraction of a larger field-of-view (FOV) block from cascade methods. \\
Our main contributions can be briefed as follows:
\begin{itemize}
  \item Introduce a new, more robust, and efficient architecture using Unet and TransUnet networks.  
  \item Add a WASP-KC block for the TransUnet model after the CNN block.
\end{itemize}
Through hands-on experiments on the 2018 Data Science Bowl challenge dataset, the results showed that the proposed network has achieved fairly good accuracy compared to other SOTA architectures on this same data set. Specifically, we have obtained 2 parameters DSC and IoU with values of 0.9225 and 0.8613.\\
The following is the organization of this paper: Firstly, the related work is described in section II. Section III introduces our proposed model. Experimental results on the 2018 Data Science Bowl challenge dataset is obtained in Section IV. Finally, the summaries, limitations, and further work are described in Section V.
\section{Related Work}

\subsection{ (Unet)}
Unet was first proposed in 2015, known as an effective Convolutional Network for Biomedical Image Segmentation. Unet architecture contains two paths, an encoder and a decoder. The encoder path is the downsampling part, each block has a rectified linear unit (ReLU) and a 2x2 max pooling operation with stride 2 \cite{b9}. The decoder path is the upsampling part for reconstructing the high-resolution feature map of the image. In particular, Unet uses skip connections to preserve spatial information because during downsampling at the encoder path, spatial information of the input image is lost, causing architecture accuracy degradation. \\

\subsection{ (ViT)}

In terms of natural language processing tasks (NLP), it has been known that Transformer architecture is one of the key criteria. However, when it comes to Computer Vision tasks, this model still has many limitations \cite{b3}. Vision transformer (ViT) is a pioneering model that adapts the transformer model to Computer Vision (CV) tasks by embedding input images into a series of visual tokens and modeling the global dependencies among this sequence with a group of transformer blocks. ViT simply processes the input image as a 1D sequence which leads to a lack of inductive bias in modeling local visual structures \cite{b11}. Recently, ViT has achieved highly competitive accuracy benchmarks in a variety of applications: image classification, object detection, and semantic image segmentation. Taking inspiration from processing input images as a sequence, the Vision transformer is a combination of the Transformer architecture part and MLP (Multilayer Perceptron) blocks \cite{b1}. The Transformer encoder of ViT includes Multi-Head Self Attention Layer (MHSA), Multi-Layer Perceptrons (MLP) Layer, and Layer Norm (LN) \cite{b12}. MHSA is the key component of the Transformer block. It is achieved after repeating single-head self-attention (SHSA) for n times, where n is the quantity of heads. MHSA is intended to reproduce long-range structural data from the images \cite{b13}.

\subsection{ (TransUnet)}
TransUnet can also be considered an upgraded version of Unet. TransUnet is the first architecture to use transformers for tasks related to Computer Vision and it has opened up new research directions with the successful application of transformers to image tasks. The big difference between TransUnet and Unet lies in the Encoder Path. There is a fairly detailed description of the TransUnet Encoder path architecture in Fig. 2. It includes CNN Block (in the study \cite{b2} the author used the backbone as ResNet50) and Vision TransFormer (ViT). The encoder which applies the transformer in ViT comprises successive layers of multiheaded self-attention (MHSA), and MPL blocks. Instead of using BatchNorm (BN), the transformer block uses LayerNorm (LN) before each one, and after each block, residual connections are put \cite{b16}\cite{b17}.

\subsection{DeepLabv3+}

In research \cite{b14}, the Atrous Spatial Pyramid Pooling module (ASPP) was proposed to be integrated with the encoder-decoder structure and this research showed better improvements to the boundaries of segmented objects in the input images. The special structure of ASPP assembled dilated convolutions in four parallel branches with distinct levels. Ultimately, being combined by fast bilinear interpolation with an additional factor of eight, the resulting feature maps were recovered to the original solution \cite{b3}. The DeepLabv3+ significantly improved over the previous version in terms of accuracy.

\subsection{Waterfall Atrous Spatial Pooling (WASP) }
The WASP is a highly efficient architecture for semantic segmentation. It leverages progressive filtering in a cascading architecture while preserving multiscale fields-of-view (FOV) in comparison with spatial pyramid configurations. According to the study in \cite {b3}, WASP when combined with the Resnet backbone will provide a robust architecture and obtain potential results for segmentation problems. Furthermore, this variation has effective computation, which is an Atrous Spatial Pooling (ASP) class variant in the DeepLabv3+ architecture. \cite{b15} demonstrated the great improvement of the WASP module in terms of computation time in training progress and decreasing parameters compared to the original ASPP module.

\section{Methodology}

\subsection{Waterfall Atrous Spatial Pooling with Skip Connection (WASP-KC) Module}\label{AA}

\begin{figure}[htbp]
\centerline{\includegraphics[width=8.9cm,height=6.3cm]{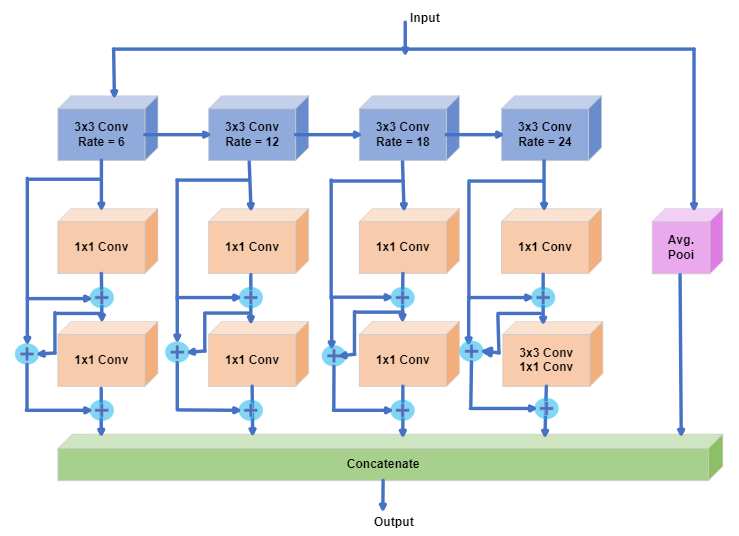}}
\caption{Waterfall Atrous Spatial Pooling (WASP-KC) module}
\label{fig}
\end{figure}

The WASP-KC module, shown in Fig.1, is inspired by the WASP module. The WASP-KC involves four units of a large-FOV that merges together and create a waterfall shape to give output.% WASP module has a strong architecture with Atrous Convolutions that is capable of leveraging both the larger FOV of the ASPP configuration and downsizing the cascade approach \cite{b3}. Unlike ASPP block, the WASP shares sequential parameters to subsequent branches, thereby extracting more information and more branch learning correlations with each other.%% đang nói WASP-KC sao lại đưa nhiều về WASP chỗ này?
Besides multiscale approaches \cite{b26}\cite{b23}, this module is also inspired by the cascade configuration \cite{b3}\cite{b14}, as well as by the parallel structures of ASPP \cite{b24} and Res2Net modules \cite{b25}.The WASP module helps to reduce parameters and memory required, which leads to less expensive computation,the main limitation of Atrous Convolutions \cite{b3}\cite{b15}. According to the experiments performed by the authors in \cite{b3}, the WASP module successfully reduced 20.69\% of the parameters and also increases the model's performance by 2\% (mIoU) using WASPnet network built on this module compared to the Res2Net-Seg or ASPP modules. In this research,  we have replaced the WASP block with the WASP-KC block by Dense connections, which are inspired by the DenseNet model. In this technique, every single layer takes all previous layers' output as input, and its feature map will be brought to deeper layers, which means that each layer receives the whole information from the previous ones. This will ensure feature reusability since feature maps of prior layers are held and added altogether which helps input image data be well kept without any loss. 
This is a significant modification that makes WASP can function more robustly. The WASP-KC block is added right after the CNN module (ResNet-50 backbone is used) to improve the performance and efficiency of the proposed model.
\begin{figure*}
  \includegraphics[width=\textwidth,height=10.1cm]{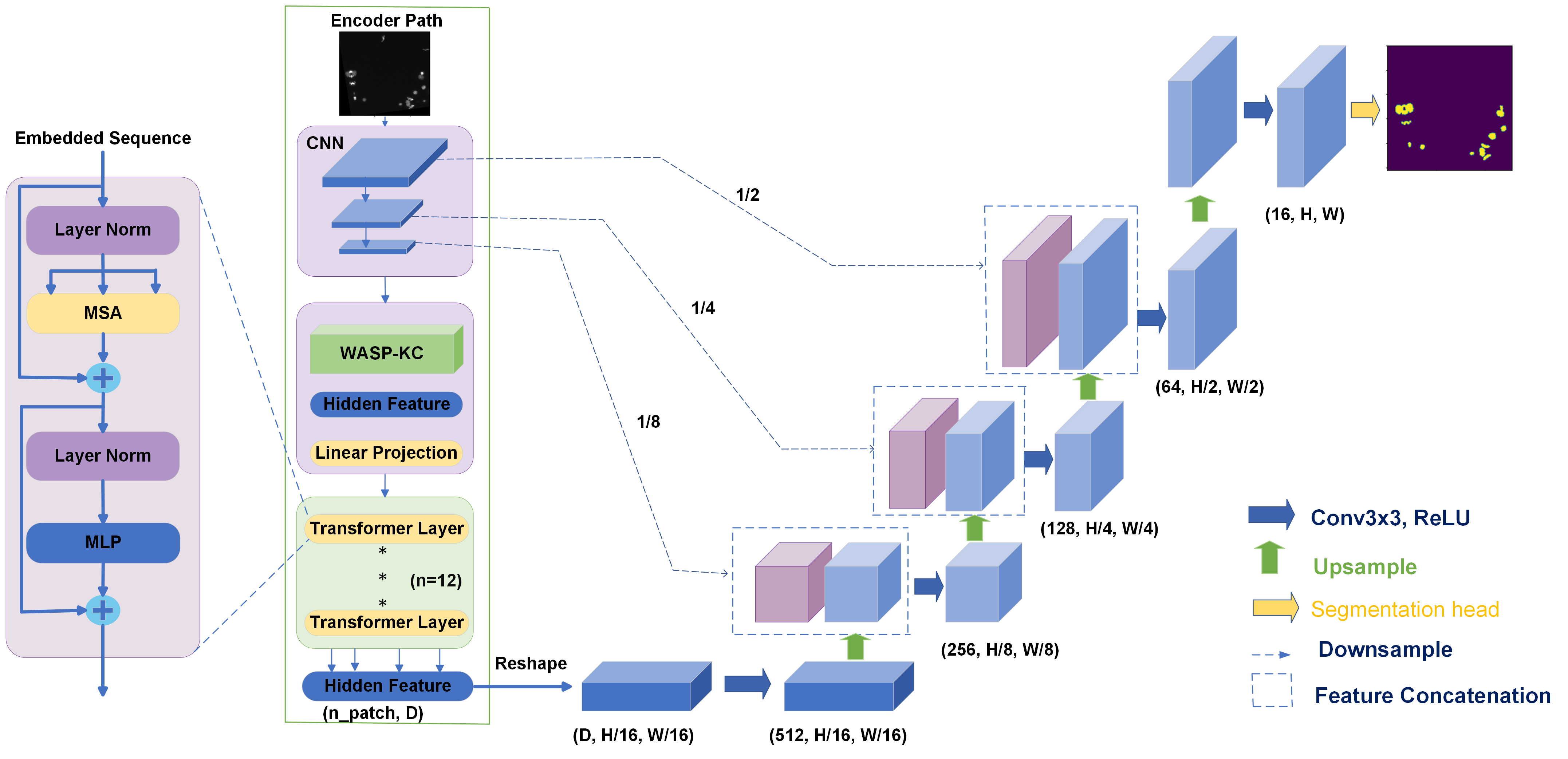}
  \caption{Architecture of TransUnet after adding WASP-KC module}
\end{figure*}

\begin{figure}[htbp]
\centerline{\includegraphics[width=8.9cm,height=3.5cm]{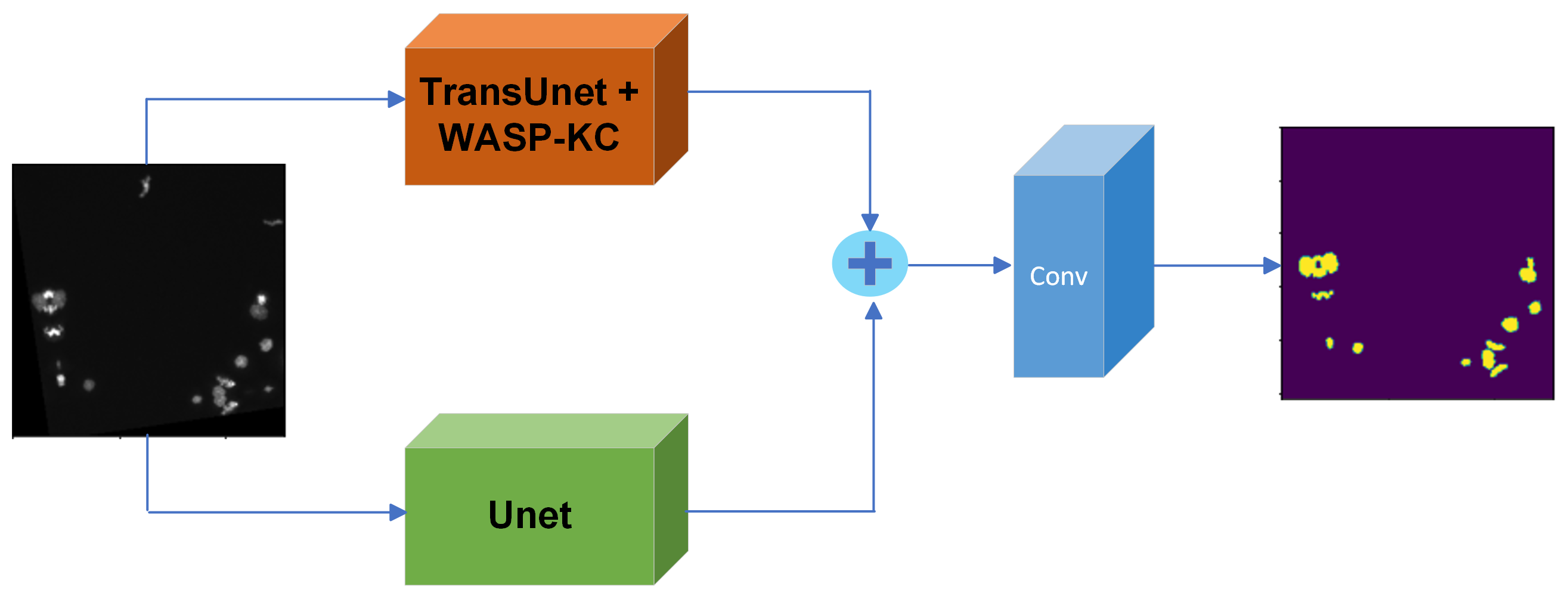}}
\caption{Architecture of Trans2Unet}
\label{fig}
\end{figure}

\subsection{Model architecture}
Aiming at developing a new deep learning architecture for nuclei cell image segmentation, this study proposed the Trans2Unet which combines Unet and TransUnet branches. First, to increase the efficiency of the TransUnet branch, we used an additional the WASP-KC block as shown in Fig. 2. The WASP-KC block consists of four convolution units. Each unit includes of three blocks, the first block uses convolution 3x3, followed by two blocks applying convolution  1x1. The 3x3 convolution blocks share information horizontally, through which the information will be used in all units of the module. In addition, skip connection is used in each unit to use the features of the previous layers. This adjustment has improved performance significantly compared to the WASP module. The output of the module is the sum of these 4 units and output of the global average pooling block, and will also be the input of the ViT network. \\
%Our research developed a new deep learning architecture for nuclei cell image segmentation. The proposed Trans2Unet which combines Unet and TransUnet branches %is demonstrated in Fig. 3. In the TransUnet branch, the WASP-KC module, which combined skip connections and the cascaded approach used in \cite{b16}, was added. The module that used the cascaded approach obtain a wider field-of-view from traditional ASPP in DeepLab for the deconvolutional stages of semantic segmentation. 

Fig. 3 shows the general structure of the proposed Trans2Unet that includes the Unet branch and the proposed TransUnet+WASP-KC branch. After the input image has been forwarded through these two branches, the outputs of the two branches will be concatenated together. And finally, after aggregating the output of the two branches above, we continue to forward through a Convolution block before making the predicted output. This is a fairly new and simple combination, but it improves performance much better than just using Unet or TransUnet as usual.\\

\subsection{Loss function}
The loss function, also known as the cost function, is an equation representing the relationship between q (which is the model's predicted result) and p (which is the actual value). Our task is to minimize the value of this equation. The loss function is used to optimize models and this is also one of the parameters to evaluate the quality of the model. Tasks related to image segmentation have many loss functions applied such as Binary Cross-Entropy (BCE), Dice loss, ... \\
\textbf{The binary cross-entropy (BCE) loss function} calculates the difference between two probability distributions, they are the actual probability distribution p and the predicted probability distribution q. It is commonly used for object classification tasks, and in image segmentation tasks as it is classification on pixels. This should be used for balanced datasets. BCE loss is represented by the following equation:

\begin{equation}
L(p, q) = -y\log(q) - (1 - p)\log(1-q)
\end{equation}

Where p represents the ground truth label, q represents the predicted value of the Trans2Unet model. The value of (1) reflects the difference between the actual value and the value predicted from the model. 

\textbf{Dice Loss} is a loss function that is popularly used in tasks relating to arcing image segmentation or medical image segmentation. . The value of this loss function measures the difference between the ground truth and the predicted value. Dice Loss is represented by the following equation:
\begin{equation}
DL(p, q) = 1 - \frac{2pq + 1} 
{p + q + 1}
\end{equation}
Mathematical notations $(p, q)$ have the meaning similar to Binary Cross-Entropy part.

\subsection{Evaluation Metrics}
Currently, The Dice Similarity Score (DSC) and Jaccard Index or Intersection over Union (IoU) are the most popular indexes for evaluating models in medical image segmentation \cite{b18}\cite{b19}\cite{b20}. In this research, we also use these two parameters to make a fair comparison with other models on the 2018 Data Science Bowl challenge dataset. 
\\ The DCS and IoU are defined by the following mathematical expressions
 \cite{b21}:
\begin{equation}
DSC =\frac{2TP} 
{2TP+FP+FN}
\end{equation}

Where: TP, FP, FN, TN are the number of true positive, false positive, false negative, and negative predictions. In addition, in the study\cite{b22}, there are other evaluation metrics for task image segmentation such as Precision, Accuracy, Volumetric Similarity, ...

\begin{equation}
IoU =\frac{TP} 
{TP+FP+FN}
\end{equation}

\section{EXPERIMENTAL RESULTS}

\subsection{Dataset}\label{AA}
To properly assess the performance evaluation of Trans2Unet model, we used the public biomedical image dataset  - the 2018 Data Science Bowl challenge dataset and GlaS dataset. The 2018 Data Science Bowl challenge dataset contains the original images, along with their masks (or ground-truth). There are 670 images in total, we splitted this dataset into the ratio of 80\% - 10\% - 10\% corresponding to the training set - validation set - test set. Some of State-of-the-art models tested on 2018 Data Science Bowl such as SSFormer-L, MSRFNet, DoubleUnet, Unet++… have achieved remarkable results. Following this dataset with the same split ratio, through trials and errors, we are confident that 670 images are enough for proposed model to perform robustly. GlaS dataset contains 165 microscopic images and
the corresponding target mask annotations. In this work we split GlaS dataset into 85 training images and 80 testing images. 
\subsection{Implementation detail}\label{AA}
We have implemented this entire proposed architecture with the Pytorch framework and conducted experiments with NVIDIA K80 GPUs. The Adam optimization function has been deprecated, with the initial learning rate (LR) set to 0.0003, and we also used a dropout regularization with p = 0.2. After three epochs with no improvement, the new learning rate is calculated by multiplying the current learning rate by a factor, which is a small value enough to reduce current learning rate and global minimum is still reached. All images in the 2018 Data Science Bowl challenge and GlaS dataset will be resized to 256 x 256 resolution. Batch size used is 10 and the number of epochs to train our model is 300.

\subsection{Evaluation}\label{AA}

In this research, we referred our model to some of the models that achieved remarkable results in the 2018 Data Science Bowl challenge and GlaS dataset to objectively review the effectiveness of this approach. In table 1, the scores reported by previous algorithms in terms of average values of the Dice Similarity Score (DSC) and IoU indexes are compared with those by the proposed approach. The table shows that our new approach gives the results that are confirmed to be good on the 2018 Data Science Bowl challenge dataset with the values of DSC - IoU are 0.9225, and 0.8613 respectively (when we fused Unet with TransUnet).

As described in Table 1, the number of Trans2Unet parameters are up to 110M, which is a disadvantage that needs improvement in upcoming research, whereas those of SSFormer-L model are 66.2M. The explanation for huge size of our proposed network is due to ViT model used in TransUnet branch. As \cite{b1}, there are 3 variants of Vit consisting of ViT-Base (86M parameters), ViT-Large (307M parameters), and ViT-Huge(632M parameters). Considering these sizes, we decided to use ViT-Base model in our network.

\begin{table}[http]
\centering
 \begin{tabular}{c | c | c | c} 
 \hline
 \textbf{Method} & \textbf{Dice Coefficient} & \textbf{Mean IoU} & \textbf{Parameters (M)} \\ [0.4ex] 
 \hline
 SSFormer-L \cite{b27}     &    \textbf{0.9230}   &       \textbf{0.8614} & \textbf{66.2} \\ 
 TransUnet    &    0.9027   &     0.8413  & 105.9\\
 MSRF-Net \cite{b28}     &    0.9224   &       0.8534 & 18.38 \\
 FANet \cite{b29}    &    0.9176   &       0.8569 & 5.76 \\
 DoubleUNet \cite{b30}     &    0.9133      &       0.8407 & 29.29 \\
 Trans2Unet (Ours)   &    0.9225   &       0.8613 & 110 \\
%  Trans2Unet (Attention Unet + TransUnet)    &    \textbf{0.9255}     &      \textbf{0.8619} \\ [1ex]
 \hline
 \end{tabular}
 \linebreak
 \caption{Performances comparison of various model on the 2018 Data Science Bowl challenge dataset.}
\end{table}
Although our results are still modest compared to other current SOTA architectures on this dataset, we believe that with this approach, the architecture will be improved in the future.

To show the improvement of the Trans2Unet model integrating with the WASP-KC module more clearly, IoU and Dice metrics of this model were compared with those of the original TransUnet as well as the Trans2Unet model integrating with the original WASP, and all experiments were tested on the same device.
The results reported in table II show that IoU and Dice metrics of our proposed model are second to none, specifically, this model has the IoU and Dice metrics are 86.13\% and 92.25\%, respectively.

\begin{table}[http]
\centering
 \begin{tabular}{c | c | c} 
 \hline
 \textbf{Method} & \textbf{Dice Coefficient} & \textbf{Mean IoU} \\ [0.4ex] 
 \hline
 TransUnet    &    0.9027   &     0.8413  \\
 Trans2Unet + WASP    &    0.9150   &     0.8499  \\
 Trans2Unet + WASP-KC   &    \textbf{0.9225}   &       \textbf{0.8613} \\
%  Trans2Unet (Attention Unet + TransUnet)    &    \textbf{0.9255}     &      \textbf{0.8619} \\ [1ex]
 \hline
 \end{tabular}
 \linebreak
 \caption{Performances comparison of Trans2Unet with WASP-KC module and its baseline models.}
\end{table}

As can be seen from Table 3, the results show that proposed network Trans2Unet also obtained great performance on GlaS dataset with Dice Coefficient of 89.94\% and Mean IoU of 82.54\%. 

\begin{table}[http]
\centering
 \begin{tabular}{c | c | c} 
 \hline
 \textbf{Method} & \textbf{Dice Coefficient} & \textbf{Mean IoU} \\ [0.4ex] 
 \hline
 FCN\cite{b32}    &    0.6661   &     0.5058  \\
 Unet\cite{b9}    &    0.7778   &     0.6534  \\
 Res-Unet\cite{b33}    &    0.7883   &     0.6595  \\
 Axial Attention Unet\cite{b34}    &    0.7630   &     0.6303  \\
 KiU-Net\cite{b35}    &    0.8325  &     0.7278  \\
 Trans2Unet (Ours)   &    \textbf{0.8984}   &       \textbf{0.8254} \\
%  Trans2Unet (Attention Unet + TransUnet)    &    \textbf{0.9255}     &      \textbf{0.8619} \\ [1ex]
 \hline
 \end{tabular}
 \linebreak
 \caption{Comparisons with various method on GlaS Dataset.}
\end{table}

\subsection{Results}\label{AA}
To demonstrate the performance of the new architecture on the 2018 Data Science Bowl challenge dataset, we show the learning curves in Fig. 4. As shown in this figure, the  model loss and scores including the Dice (DSC), and IoU converge after 100 epochs and stay stable. For qualitative assessment, we also show some representative segmentation results of the test set of this dataset in Fig. 5. It is obvious in Fig.5, the predictions by the proposed approach are in good agreement with those by ground truths.

\begin{figure}[htbp]
\centerline{\includegraphics[width=8.9cm,height=3cm]{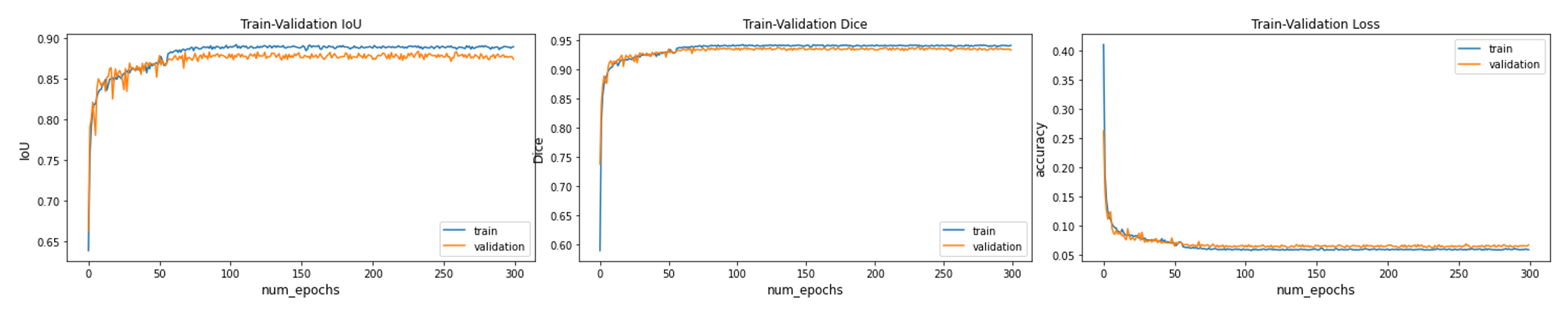}}
\caption{Training curves on the 2018 Data Science Bowl challenge dataset}
\label{fig}
\end{figure}

\begin{figure}[htbp]
\centerline{\includegraphics[width=7.5cm,height=10.5cm]{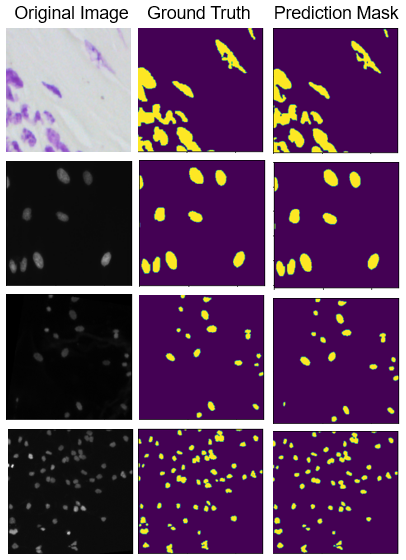}}
\caption{Some representative segmentation results of Trans2Unet on Nuclei images from 2018 Data Science Bowl challenge dataset}
\label{fig}
\end{figure}

\section{Conclusion}
In this study, we have introduced a new architecture, which is a combination of two other deep learning networks, Unet and TransUnet, for nuclei image segmentation. Furthermore, to approach leverages the progressive extraction of larger fields-of-view (FOV) from cascade methods, we integrated WASP-KC (WASP module with Skip Connections) module into the TransUnet architecture. Through experiments on the 2018 Data Science Bowl challenge dataset, we show that our proposed model has achieved quite good results expressed through DSC or IoU scores. By combining the Unet with the TransUnet architecture, the model can maintain local features of CNN and take advantage of global features in Transformers for more robust segmentation. We believe that this structure can be a good approach to improve the efficiency of models not only for nuclei cell but also for general tasks of image segmentation.
%In addition, we hope this approach of combining deep learning models will achieve new, more powerful, more efficient architectures, and it will be a potential idea to be able to continue to develop more robust models in the future.%

\section*{Acknowledgment}
This research is funded by the Hanoi University of Science and Technology (HUST) under project number T2021-
PC-005.

\vspace{12pt}


\begin{thebibliography}{00}
\bibitem{b1} Dosovitskiy, A., Beyer, L., Kolesnikov, A., Weissenborn, D., Zhai, X., Unterthiner, T., Dehghani, M., Minderer, M., Heigold, G., Gelly, S., et al.: An image is worth 16x16 words: Transformers for image recognition at scale. In: ICLR (2021)

\bibitem{b2} Chen, Jieneng and Lu, Yongyi and Yu, Qihang and Luo, Xiangde and Adeli, Ehsan and Wang, Yan and Lu, Le and Yuille, Alan L., and Zhou, Yuyin. TransUNet: Transformers Make Strong Encoders for Medical Image Segmentation. arXiv preprint arXiv:2102.04306, 2021
\bibitem{b3} Bruno Artacho and Andreas E. Savakis. Waterfall atrous spatial pooling architecture for efficient
semantic segmentation. CoRR, abs/1912.03183, 2019.
\bibitem{b4} 2018 Data Science Bowl, https://www.kaggle.com/c/data-science-bowl-2018/overview
\bibitem{b5} M. E. Celebi, N. Codella, and A. Halpern, “Dermoscopy image analysis:
overview and future directions,” IEEE J. Biomed. Health Inform, vol.
23, no. 2, pp. 474–478, 2019.
\bibitem{b6} J. C. Caicedo et al., “Nucleus segmentation across imaging experiments:
the 2018 data science bowl,” Nat. Meth., vol. 16, no. 12, pp. 1247–1253, 2019.
\bibitem{b7} S. Ali et al., “Deep learning for detection and segmentation of artefact
and disease instances in gastrointestinal endoscopy,” Med. Imag. Anal., p. 102002, 2021.
\bibitem{b8} Caicedo, J. C. et al. Evaluation of deep learning strategies for nucleus
segmentation in fluorescence images. Cytom. A 95, 952–965 (2019).
\bibitem{b9} Ronneberger, O., Fischer, P., Brox, T.: U-net: Convolutional networks
for biomedical image segmentation. In: International Conference on
Medical image computing and computer-assisted intervention. pp. 234- 241. Springer (2015)
\bibitem{b10} Chen, D., Yang, W., Wang, L., Tan, S., Lin, J., Bu, W. (2022). PCATUNet: UNet-like network fused convolution and transformer for retinal vessel segmentation. PLoS ONE, 17(1), e0262689.
\bibitem{b11} Zhang, Qiming and Xu, Yufei and Zhang, Jing and Tao, Dacheng.
ViTAEv2: Vision Transformer Advanced by Exploring Inductive Bias
for Image Recognition and Beyond. arXiv preprint arXiv:2202.10108, 2022
\bibitem{b12} Ashish Vaswani, Noam Shazeer, Niki Parmar, Jakob Uszkoreit, Llion
Jones, Aidan N Gomez, Łukasz Kaiser, and Illia Polosukhin. Attention
is all you need. In NeurIPS, 2017.
\bibitem{b13} Petit, O., Thome, N., Rambour, C., Themyr, L., Collins, T., Soler,
L. (2021). U-Net Transformer: Self and Cross Attention for Medical Image Segmentation. In: Lian, C., Cao, X., Rekik, I., Xu, X.,
Yan, P. (eds) Machine Learning in Medical Imaging. MLMI 2021.
Lecture Notes in Computer Science(), vol 12966. Springer, Cham.
https://doi.org/10.1007/978-3-030-87589-32 8
\bibitem{b14} Badrinarayanan, V.; Kendall, A.; Cipolla, R. SegNet: A Deep Convolutional Encoder-Decoder Architecture for Image Segmentation. arXiv 2015, arXiv:1511.00561
\bibitem{b15} Sharma, Shorya. Semantic Segmentation for Urban-Scene Images, 2021
\bibitem{b16} Alexei Baevski and Michael Auli. Adaptive input representations for neural language modeling. In
ICLR, 2019.
\bibitem{b17} Qiang Wang, Bei Li, Tong Xiao, Jingbo Zhu, Changliang Li, Derek F. Wong, and Lidia S. Chao.
Learning deep transformer models for machine translation. In ACL, 2019.
\bibitem{b18} Kamnitsas, K., Ledig, C., Newcombe, V.F., Simpson, J.P., Kane, A.D.,
Menon, D.K., Rueckert, D., Glocker, B.: Efficient multi-scale 3D CNN
with fully connected CRF for accurate brain lesion segmentation. MIA
36, 61–78 (2017)
\bibitem{b19} Ronneberger, O., Fischer, P., Brox, T.: U-net: Convolutional networks
for biomedical image segmentation. MICCAI pp. 234–241 (2015)
\bibitem{b20} BRATS challenge (2018), https://www.med.upenn.edu/sbia/brats2018.html
\bibitem{b21} Taha AA, Hanbury A. Metrics for evaluating 3D medical image
segmentation: Analysis, selection, and tool. BMC Med Imaging
[Internet]. 2015 Aug 12 [cited 2021 May 14];15(1):29. Available from:
http://bmcmedimaging.biomedcentral.com/articles/10.1186/s12880-015- 0068-x
\bibitem{b22} Muller, Dominik and Hartmann, Dennis and Meyer, Philip and Auer, ¨
Florian and Soto-Rey, Inaki and Kramer, Frank. MISeval: a Metric ˜
Library for Medical Image Segmentation Evaluation. arXiv, 2022.
\bibitem{b23} Roy, A.; Todorovic, S. A Multi-Scale CNN for Affordance Segmentation
in RGB Images. In Proceedings of the IEEE European Conference on
Computer Vision (ECCV), Amsterdam, the Netherlands, 11–14 October
2016; pp. 186–201.
\bibitem{b24} Chen, L.C.; Papandreou, G.; Kokkinos, I.; Murphy, K.; Yuille, L.
DeepLab: Semantic Image Segmentation with Deep Convolutional Nets,
Atrous Convolution and Fully Connected CFRs. IEEE Trans. Pattern
Anal. Mach. Intell. 2018, 40, 834–845.
\bibitem{b25} Gao, S.H.; Cheng, M.M.; Zhao, K.; Zhang, X.Y.; Yang, M.H.; Torr,
P. Res2Net: A New Multi-Scale Backbone Architecture. IEEE Trans.
Pattern Anal. Mach. Intell. 2019.
\bibitem{b26} Eigen, D.; Fergus, R. Predicting Depth, Surface Normals and Semantic
Labels with a Common Multi-Scale Convolutional Architecture. arXiv
2014, arXiv:1411.4734.
\bibitem{b27} Wang, Jinfeng and Huang, Qiming and Tang, Feilong and Meng, Jia and Su, Jionglong and Song, Sifan. Stepwise Feature Fusion: Local Guides Global, 2022.
\bibitem{b28} Srivastava, A., Jha, D., Chanda, S., Pal, U., Johansen, H.D., Johansen, D., Riegler,
M.A., Ali, S., Halvorsen, P.: MSRF-Net: A Multi-scale Residual Fusion Network for
Biomedical Image Segmentation. IEEE Journal of biomedical imaging and health
informatics (2021)
\bibitem{b29} Tomar, Nikhil Kumar, et al. "Fanet: A feedback attention network for improved biomedical image segmentation." IEEE Transactions on Neural Networks and Learning Systems 34.11 (2022): 9375-9388.
\bibitem{b30} D. Jha et al., “Doubleu-net: A deep convolutional neural network for medical image segmentation,” 07 2020, pp. 558–564.
\bibitem{b31} Zongwei Zhou, Md Mahfuzur Rahman Siddiquee, Nima Tajbakhsh, and Jianming Liang. Unet++: A nested u-net architecture for medical image segmentation. In zhou2018unet, pages 3–11. 2018.
\bibitem{b32} Long, J., Shelhamer, E. \& Darrell, T. Fully convolutional networks for semantic segmentation. Proceedings Of The IEEE Conference On Computer Vision And Pattern
Recognition. pp. 3431-3440 (2015)
\bibitem{b33} Jha, D., Smedsrud, P., Riegler, M., Johansen, D., De Lange, T., Halvorsen, P.
\& Johansen, H. Resunet++: An advanced architecture for medical image segmentation. 2019 IEEE International Symposium On Multimedia (ISM). pp. 225-2255 (2019)
\bibitem{b34} Valanarasu, V. (2021). Medical Transformer: Gated Axial-Attention for Medical Image Segmentation. In Medical Image Computing and Computer Assisted Intervention – MICCAI 2021 (pp. 36–46). Springer International Publishing.
\bibitem{b35} Valanarasu, J., Sindagi, V., Hacihaliloglu, I. \& Patel, V. Kiu-net: Towards accurate segmentation of biomedical images using over-complete representations. International Conference On Medical Image Computing And Computer-Assisted Intervention. pp. 363-373 (2020)

\end{thebibliography}
\end{document}